\documentclass[aps,pr,superscriptaddress,preprintnumbers,nofootinbib,10pt]{revtex4-2}

\usepackage{multirow}
\usepackage{amsmath}
\usepackage{amssymb}
\usepackage[dvipdf,dvips]{graphicx}
\usepackage{color}
\usepackage{hyperref}
\usepackage{url}
\usepackage{slashed}
\usepackage{subfigure}
\usepackage[usenames,dvipsnames]{xcolor}
\usepackage{amsmath}
\usepackage{amsfonts}
\usepackage{float} 
\usepackage{amssymb}
\usepackage{epsfig}
\usepackage{graphics}
\usepackage{euscript}
\usepackage{slashed}
\usepackage{epstopdf}
\usepackage[utf8]{inputenc}
\allowdisplaybreaks
\usepackage[normalem]{ulem}
\usepackage{pifont}
\usepackage{dsfont}
\usepackage{MnSymbol}
\usepackage{verbatim}
\usepackage{graphicx}
\usepackage{latexsym}
\usepackage{tikz-feynman}
\usepackage{bbold}

\begin{document}

	\title{{\bf \Large Introduction to Bell's inequality in Quantum Mechanics}}
	
	\vspace{1cm}
	
	\author{M. S.  Guimaraes}\email{msguimaraes@uerj.br} \affiliation{UERJ $–$ Universidade do Estado do Rio de Janeiro,	Instituto de Física $–$ Departamento de Física Teórica $–$ Rua São Francisco Xavier 524, 20550-013, Maracanã, Rio de Janeiro, Brazil}
	
	\author{I. Roditi} \email{roditi@cbpf.br} \affiliation{CBPF $-$ Centro Brasileiro de Pesquisas Físicas, Rua Dr. Xavier Sigaud 150, 22290-180, Rio de Janeiro, Brazil}
	
	\author{S. P. Sorella} \email{silvio.sorella@fis.uerj.br} \affiliation{UERJ $–$ Universidade do Estado do Rio de Janeiro,	Instituto de Física $–$ Departamento de Física Teórica $–$ Rua São Francisco Xavier 524, 20550-013, Maracanã, Rio de Janeiro, Brazil}

	\footnote{Lectures given at the Ph.D programs of the Physics Institute of the State University of Rio de Janeiro, UERJ, and of the CBPF.}

	\begin{abstract}
	A pedagogical introduction to Bell's inequality in Quantum Mechanics is presented. Several examples, ranging from spin $1/2$ to coherent and squeezed states are worked out. The generalization to Mermin's inequalities and to GHZ states is also outlined.  

	\end{abstract}

	\maketitle

	\vspace{1cm}
	
	\tableofcontents

	\section{Introduction}\label{Intr} 
	
	The Bell inequality \cite{Bell64} is a milestone of Quantum Mechanics. It highlights in a surprising way the existence of  strong correlations between the components of a composite system. These correlations cannot be reproduced by any local hidden variable theory, surviving over very long distances. They are the signature of one of the most fascinating phenomenon of the quantum world: {\it the entanglement}. \\\\The experimental confirmation of the violation of the Bell inequality by the pioneering works of A. Aspect, J. Clauser and A. Zeilinger \cite{Claus1,Claus2, Claus3, Claus4, Asp1, Asp2, Asp3,Asp4, Z0,Z1}
gives to Quantum Mechanics the status of the unique theory able to consistently facing the intricacies and the complexity of the entanglement.\\\\The whole story goes back to the famous  paper by Einstein-Podolsky-Rosen \cite{EPR}. This work is considered  the birth of the study of the entanglement. The phenomenon was so challenging that led Einstein-Podolsky-Rosen to argue that Quantum Mechanics was an incomplete theory. This statement marked the beginning of a long and difficult debate on the foundations of the Quantum Theory. \\\\It was only in 1964, the year of the publication of Bell's work \cite{Bell64}, that the debate was elucidated. Bell was able to derive a certain inequality whose violation would imply that Quantum Mechanics could not be reconciled with any other local hidden variable formulation. \\\\Bell's inequality is at the origin of the great achievements which have been obtained so far in many areas: Quantum Computation, Quantum Teleportation, Quantum Cryptography, etc.\\\\The purpose of these notes is that of presenting, in an introductory way, the general features of a version of the Bell inequality called the Bell-CHSH inequality \cite{CHSH69}. Many examples will be worked out, ranging from spin $1/2$ states to coherent and squeezed states. The generalization of the Bell-CHSH inequality to multipartite systems, known as the Mermin inequalities, will be also outlined. \\\\The material is organized in a self-contained  way. Students with a basic knowledge of Quantum Mechanics will be able to reproduce all calculations and solve the proposed exercises. 
	
	\section{Entangled and product states. Spin $1/2$ states}
	
	The concept of entangled states can be illustrated in a simple way by making use of spin $1/2$ states. Let us begin by considering a bipartite system, $AB$, made up by two spins $1/2$. The Hilbert space of such a system is: ${\cal H}= {\cal H_A}\otimes {\cal H_B}$, where ${\cal H}_A$ and $\cal H_B$ are the Hilbert spaces of $A$ and $B$, respectively.  It is customary to refer to the capital letters $(A,B)$ as Alice and Bob (two agents measuring the system in their respective Labs). An orthonormal basis of  ${\cal H}= {\cal H_A}\otimes {\cal H_B}$ contains four elements, namely: $(|+\rangle_A\otimes |+\rangle_B,\; |+\rangle_A\otimes |-\rangle_B, \;|-\rangle_A\otimes |+\rangle_B, \;|-\rangle_A\otimes |-\rangle_B)$\footnote{As usual, the states $(|+\rangle, |-\rangle)$ are eigenstates of the diagonal Pauli matrix $\sigma_z$ along the z-axix: 
	\begin{equation}
  {\sigma_z} =
  \left[ {\begin{array}{cc}
    1 & 0 \\
    0 & -1 \\
  \end{array} } \right] \label{sz}
\end{equation}
with
\begin{equation} 
\sigma_z |+\rangle = |+\rangle \;, \qquad \sigma_z |-\rangle = - |-\rangle \;. \label{eig}
\end{equation}	
	}. Any state $|\psi\rangle$ of $\cal H$ can be written as a linear combination of the elements of the basis:
	\begin{equation} 
	|\psi\rangle = a_1|+\rangle_A\otimes |+\rangle_B + a_2 (|+\rangle_A\otimes |-\rangle_B + a_3|-\rangle_A\otimes |+\rangle_B +a_4|-\rangle_A\otimes |-\rangle_B \;, \label{exp}
	\end{equation}
where, from the normalization condition, $\langle \psi |\psi\rangle =1 $, the coefficients $(a_1,a_2,a_3,a_4)$ fulfill the constraint
\begin{equation} 
|a_1|^2 + |a_2|^2 +|a_3|^2 +|a_4|^2  = 1 \;. \label{norm}
\end{equation}	
 The state $|\psi\rangle$ is said to be a {\it product} state if it can be written as the tensor product of  states belonging to  $\cal H_A$ and  ${\cal H}_B$, {\it i.e.}
 \begin{equation}
 |\psi\rangle = |\psi\rangle_A \otimes |\psi \rangle_B \;, \label{pd}
 \end{equation}
	where $(|\psi\rangle_A, |\psi \rangle_B)$ stand for vectors of ${\cal H_A}$, ${\cal H_B}$: 
	\begin{equation} 
	|\psi\rangle_A  = (c_1 |+\rangle_A + c_2 |-\rangle_A) \;, \qquad 	|\psi\rangle_B  = (b_1 |+\rangle_B + b_2 |-\rangle_B) \;. \label{psiab}
	\end{equation} 
One sees thus that $|\psi\rangle$ in eq.\eqref{exp} is a product stste if $(a_1,a_2,a_3,a_4)$ can be written as 
\begin{equation}
a_1 = c_1 b_1\;, \qquad a_2 = c_1 b_2\;, \qquad a_3 = c_2 b_1\;, \qquad a_4 = c_2 b_2\;. \label{pcd}
\end{equation}	
The above condition can be linked to a $2\times 2$ determinant made up by $(a_1,a_2,a_3,a_4)$. Let us encode  the coefficients of the expansion \eqref{exp} in the matrix ${\cal A} $
\begin{equation}
  {\cal A} =
  \left[ {\begin{array}{cc}
    a_{1} & a_{2} \\
    a_{3} & a_{4} \\
  \end{array} } \right] \label{mt}
\end{equation}
Conditions \eqref{pcd} imply that the determinant of ${\cal A}$ vanishes: 
\begin{equation}
{\rm det}{\cal A} = a_1 a_4 - a_2 a_3 = c_1b_1c_2 b_2 - c_1 b_2 c_2 b_1 = 0 \;. \label{dt}
\end{equation}
This equation gives  a quick way of checking if a given state can be written as a product state or not. Suppose that the determinant of ${\cal A}$ does not vanish, As a consequence, eq.\eqref{pcd} cannot be fulfilled, so that the state cannot be written as a product state. \\\\In such a case, {\it i.e.} when $|\psi\rangle$ cannot be written as in eq.\eqref{pcd}, one says that it is {\it entangled}: 
\begin{equation} 
|\psi\rangle \; \; {\rm entangled}\;\; \Rightarrow |\psi\rangle \neq |\psi\rangle_A \otimes |\psi \rangle_B
\end{equation} 
\noindent {\it {\bf  {\underline{Exercise}}}} 
Show that the state 
\begin{equation} 
|\psi\rangle_N = \frac{1}{\sqrt{N}} \left( |+\rangle_A \otimes |+\rangle_B + |+\rangle_A\otimes  |-\rangle_B + |-\rangle_A\otimes  |+\rangle_B \right) 
+ \frac{\sqrt{N-3}}{\sqrt{N} } |-\rangle_A \otimes |-\rangle_B  \;, \qquad N \; {\rm integer}\;, \qquad N\ge 3\;, \label{nst}
\end{equation}
	is entangled for $N\ge 5$. What happens for $N=3,4$?

	\subsection{The Bell states} \label{Bellst}
	
	Out of the four elements $(|+\rangle_A\otimes |+\rangle_B, \; (|+\rangle_A\otimes |-\rangle_B, \;|-\rangle_A\otimes |+\rangle_B, 
	\;|-\rangle_A\otimes |-\rangle_B)$ one can construct the following {\it four entangled states}  $|\phi_{\alpha}\rangle, \alpha=0,1,2,3$, called the Bell states, namely 
\begin{eqnarray} 
|\phi_{0}\rangle &=& \frac{1}{\sqrt{2}}\left( |+\rangle_A\otimes |+\rangle_B + |-\rangle_A\otimes |-\rangle_B \right)\;, \nonumber \\
|\phi_{1}\rangle &=& \frac{1}{\sqrt{2}}\left( |+\rangle_A\otimes |+\rangle_B - |-\rangle_A\otimes |-\rangle_B\right) \;, \nonumber \\
|\phi_{2}\rangle &=& \frac{1}{\sqrt{2}}\left( |+\rangle_A\otimes |-\rangle_B - |-\rangle_A\otimes |+\rangle_B\right) \;, \nonumber \\
|\phi_{3}\rangle &=& \frac{1}{\sqrt{2}}\left( |+\rangle_A\otimes |-\rangle_B + |-\rangle_A\otimes |+\rangle_B\right) \;. \label{Bs}
\end{eqnarray}
	A quick computation shows that 
	\begin{equation}
	\langle \phi_\alpha \; | \; \phi_\beta \rangle = \delta_{\alpha\beta} \;, \label{ot}
	\end{equation}
so that the Bell states form an orthonormal basis of entangled states. The state $\phi_2$ is called singlet, due to the fact that it has zero total spin:
\begin{equation}
{\vec S} |\phi_2\rangle = ({\vec S}_A + {\vec S}_B) |\phi_2\rangle = 0 \;. \label{TS}
\end{equation} 	
\vspace{1cm}
\noindent {\it {\bf  {\underline{Exercise}}}}
Prove equation \eqref{TS}. 

\noindent {\it {\bf  {\underline{Exercise}}}}\\\\Express the four elements 
	$(|+\rangle_A\otimes |+\rangle_B, \; (|+\rangle_A\otimes |-\rangle_B, \;|-\rangle_A\otimes |+\rangle_B, 
	\;|-\rangle_A\otimes |-\rangle_B)$ in terms of the Bell entangled basis.

\section{The Bell-CHSH inequality}\label{Bineq}
The meaning of the Bell-CHSH inequality can be captured by the following simplified argument\footnote{See Appendix \eqref{A} for an account of original Bell's proof.}. Alice and Bob are located in their respective laboratories, which are supposed to be far enough apart so  that Alices's measurements do not interfere with Bob's measurements and vice-versa.  In particular, as emphasized in \cite{Bell64}, the settings of Alice's device do not depend on those of Bob's device and vice-versa. The measurements carried out by both Alice and Bob have only two possible outcomes: $\pm 1$. One says that they are dichotomic measurements, {\it i.e.}  measurements, for instance, of the type yes/no, up/down, .. 
The measurements of $(A,A')$ and $(B,B')$ are repeated $n$ times. Let $(a_j, a'_j, b_j, b'_j)$ stand for the outcomes of the {\it j-th} measurement. Let us consider the combination
\begin{equation} 
\langle (A+A')B + (A-A')B'\rangle_{av} \;, \label{ccomb}
\end{equation} 
where the brackets $\langle \; \cdot \;\rangle_{av}$ mean 
\begin{equation} 
\langle (A+A')B + (A-A')B'\rangle_{av} = \frac{1}{n} \sum_{j=1}^{n} \left( (a_j+a'_j)b_j + (a_j-a'_j)b'_j \right)  \;, \label{bck}
\end{equation}
with
\begin{equation}
a_j^2 = {a'}_j^2= b_j^2 = {b'}_j^2= 1\label{pm}
\end{equation}
It holds that 
\begin{equation} 
\big| (a_j+a_j')b_j +(a_j-a_j')b_j' \big| \le 2 \;\;\;\;\forall j  \;. \label{two}
\end{equation}
In fact, making use of the triangle inequality, it follows 
\begin{equation} 
\big| (a_j+a_j')b_j +(a_j-a_j')b_j' \big| \le \big| (a_j+a_j')b_j\big| + \big|(a_j-a_j')b_j'\big| = \big|(a_j+a_j')\big| + \big|(a_j-a_j')\big|  \;. \label{two1}
\end{equation}
since $|b_j|=|b_j'|=1$. Moreover, the maximum value of $|a_j+a_j'|$ is equal to 2, corresponding to $a_j=a_j'=\pm 1$. However, when $|a_j+a_j'|=2$, one observes that $|a_j-a_j'|$ vanishes, and vice-versa, showing thus the validity of eq.\eqref{two}. 
Therefore: 
\begin{equation} 
\Big| \langle (A+A')B + (A-A')B'\rangle_{av} \Big|  \le 2  \;.  \label{bck2}
\end{equation}
\vspace{1cm}

\noindent Let us now turn to Quantum Mechanics.  Here, we deal with Hermitian dichotomic operators, $(A,A')$ and $(B,B')$, called Bell's operators. Alice's operators $(A,A')$ act on the Hilbert space $\cal H_A$, while Bob's operators $(B,B')$ on ${\cal H}_B$. These operators are required to fulfill the following conditions:
\begin{eqnarray} 
A & =& A^{\dagger} \;, \quad A'  = {A'}^{\dagger} \;, \quad B  = B^{\dagger} \;, \quad B'  = {B'}^{\dagger} \;, \nonumber \\[3mm]
A^2 & =& {A'}^2 = B^2 = {B'}^2=1 \;, \label{AB}
\end{eqnarray}
and
\begin{eqnarray}
\left[\; A,B\; \right] &= & 0\;, \quad \left[\; A',B\; \right] =  0 \;, \quad \left[\; A,B'\; \right] =  0 \;, \quad \left[\; A',B'\; \right] =  0 \;.  \nonumber \\[3mm]
\left[\; A,A'\; \right] &\neq & 0\;, \quad \left[\; B,B'\; \right] \neq  0 \;.   
\label{ABBell}
\end{eqnarray}
From equation \eqref{ABBell} one sees that Alice's operators commute with Bob's operators.  More precisely, Alice and Bob are implicitly meant to be space-like separated. As such, measurements performed by Alice do not interfere with Bob's measurements. \\\\On the other hand, $(A,A')$ have a non-vanishing commutator. The same holds for $(B,B')$. This last condition has a simple understanding. For instance, in the case of spin $1/2$, it means that Alice and Bob cannot perform two simultaneous measurements of their own spin in two different directions. \\\\In Quantum Mechanics the analogue of the combination \eqref{ccomb} is provided by the correlator 
\begin{equation} 
\langle \psi | \;{\cal C} | \;\psi \rangle \;, \label{corr}
\end{equation}
where $|\psi\rangle$ stands for the normalized state of the system, $\langle \psi|\psi \rangle=1$,  and ${\cal C}$ is given by the  Hermitian Bell-CHSH operator 
\begin{equation} 
{\cal C} = (A + A') \otimes B + (A-A')\otimes B' \;, \label{cop}
\end{equation}
The question which is addressed is whether the bound \eqref{bck2} holds at the quantum level, that is if 
\begin{equation} 
|\langle \psi | \;{\cal C} | \;\psi \rangle | \le 2  \;? \;. \label{wbbound}
\end{equation}
The answer given by Quantum Mechanics is astonishing. If the state $|\psi\rangle$ is a product state, the bound \eqref{two} still holds. However, if the state $|\psi\rangle$ is entangled, then the correlator \eqref{corr} turns out to be strictly greater than 2, namely 
\begin{equation} 
|\langle \psi | \;{\cal C} | \;\psi \rangle | _{\it entang} > 2 \;. \label{vbd}
\end{equation}
In this case, one speaks of a violation of the Bell-CHSH inequality. 
	
\subsection{Example of violation of the Bell-CHSH inequality}\label{fv}
	
As a first example of violation of the Bell-CHSH inequality, 	we consider the case in which Alice and Bob share the first Bell entangled state $|\phi_0\rangle$, eq.\eqref{Bs}. The first task is that of identifying the four Bell dichotomic operators $(A,A',B,B')$. To that end, we shall introduce the following operators 
\begin{eqnarray} 
A |+\rangle_A & = & e^{i \alpha} |-\rangle_A \;, \qquad A |-\rangle_A = e^{-i \alpha} | +\rangle_A \;, \nonumber \\
A' |+\rangle_A & = & e^{i \alpha'} |-\rangle_A \;, \qquad A' |-\rangle_A = e^{-i \alpha'} | +\rangle_A \;, \nonumber \\
B |+\rangle_B & = & e^{i \beta} |-\rangle_B \;, \qquad B |-\rangle_B = e^{-i \beta} | +\rangle_B \;, \nonumber \\
B' |+\rangle_B & = & e^{i \beta'} |-\rangle_B \;, \qquad B' |-\rangle_B = e^{-i \beta'} | +\rangle_B \;, \label{Bphi}
\end{eqnarray} 
where $(\alpha, \alpha', \beta, \beta')$ are arbitrary parameters. It is easy to show that the operators in eq.\eqref{Bphi} obey all conditions of eqs.\eqref{AB},\eqref{ABBell}. For example: 
\begin{equation}
A^2 |+\rangle_A = e^{i \alpha} A |-\rangle_A = e^{i \alpha} e^{-i \alpha} |+\rangle_A = |+\rangle_A \;. \label{one}
\end{equation}
The next step is that of evaluating $A \otimes B|\phi_0\rangle$. From eqs.\eqref{Bphi}, we get 
\begin{equation} 
A\otimes B|\phi_0\rangle = \frac{1}{\sqrt{2}}\left( e^{i(\alpha + \beta)} |-\rangle_A\otimes |-\rangle_B + e^{-i(\alpha + \beta)} |+\rangle_A\otimes |+\rangle_B \right) \;. \label{ab1}
\end{equation}
Thus
\begin{equation} 
\langle \phi_0 |\; A\otimes B \;| \phi_0 \rangle = \cos(\alpha+\beta) \;.  \label{cos}
\end{equation}	
For the Bell-CHSH correlator one obtains 
\begin{equation} 
\langle \phi_0 |\; {\cal C} \;| \phi_0 \rangle =\langle \phi_0 |\; (A+A')\otimes B+ (A-A')\otimes B' \;| \phi_0 \rangle = \cos(\alpha+\beta)+\cos(\alpha'+\beta)+\cos(\alpha+\beta')- \cos(\alpha'+\beta') \;. \label{cab1}
\end{equation} 
Expression \eqref{cab1} depends on the four arbitrary parameters $(\alpha, \alpha', \beta, \beta')$. It remains  to investigate if it exists a choice of $(\alpha, \alpha', \beta, \beta')$ which gives rise to a violation of the Bell-CHSH inequality, {\it i.e.}
$|\langle \phi_0 |\; {\cal C} \;| \phi_0 \rangle| >2$.\\\\The standard choice which one finds in textbooks \cite{ZW,AP,NC,SC} is 
\begin{equation} 
\alpha = 0 \;, \qquad \alpha'= \frac{\pi}{2} \;, \qquad \beta = - \frac{\pi}{4} \;, \qquad \beta'= \frac{\pi}{4} \;, \label{ang}
\end{equation}
yielding 
\begin{equation} 
|\langle \phi_0 |\; {\cal C} \;| \phi_0 \rangle| = 2 \sqrt{2} = 2.8284 \;. \label{fv}
\end{equation}
We see that the Bell state $|\phi_0\rangle$ yields  a violation of the Bell-CHSH inequality. As we shall see in the next Sections, the value $2\sqrt{2}$, known as Tsirelson's bound \cite{TSI}, is the maximum value attainable by the Bell-CHSH correlator $|\langle \psi | {\cal C} |\psi\rangle|$. Replacing $|\phi_0\rangle$  by any other Bell's state yields the same value for the violation: $2\sqrt{2}$. All four Bell's states violate maximally the Bell-CHSH inequality. \\\\Let us conclude this Section with a remark on the Bell operators of eq.\eqref{Bphi}. These operators are a particular case of a more general set of dichotomic operators built out with the three Pauli matrices 
\begin{equation}
  {\sigma_x} =
  \left[ {\begin{array}{cc}
    0 & 1 \\
    1 & 0 \\
    \end{array} } \right]
     \;, \qquad 
    {\sigma_y} =
  \left[ {\begin{array}{cc}
    0 & -i \\
    i & 0 \\
    \end{array} } \right]
     \;, \qquad 
    {\sigma_z} =
  \left[ {\begin{array}{cc}
    1 & 0 \\
    0 & -1 \\ 
  \end{array} } \right] \;. \label{pam}
\end{equation}
From 
\begin{equation} 
\sigma_i \; \sigma_j = \delta_{ij} + i \varepsilon_{ijk}\; \sigma_k \;, \label{sijk}
\end{equation}
it follows that the operator $({\hat n}\cdot {\vec \sigma})$, where $\hat n$ is a unit vector, is dichotomic and Hermitian 
\begin{equation} 
({\hat n}\cdot {\vec \sigma})^2 = 1 \;, \qquad ({\hat n}\cdot {\vec \sigma})^{\dagger} = ({\hat n}\cdot {\vec \sigma}) \;. \label{ds}
\end{equation}
Using polar coordinates, ${\hat n}=(\sin{\theta} \cos{\varphi}, \sin{\theta} \sin{\varphi}, \cos{\theta})$, we have 
\begin{equation} 
 {{\hat n}\cdot {\vec \sigma}} =
  \left[ {\begin{array}{cc}
    \cos{\theta} & e^{-i \varphi} \sin{\theta} \\
    e^{i \varphi} \sin{\theta} & - \cos{\theta} \\
    \end{array} } \right] \;. \label{pcoo}
\end{equation}
For the Bell operators we get the more general form 
\begin{eqnarray} 
A |+\rangle_A & = &( \cos({\theta}) |+\rangle_A + e^{i \alpha}\sin({\theta}) |-\rangle_A ) \;, \qquad \;\;\;\;\;\;A |-\rangle_A = (-\cos({\theta}) |-\rangle_A+ e^{-i \alpha}\sin({\theta}) | +\rangle_A) \;, \nonumber \\
A' |+\rangle_A & = &( \cos({\theta'}) |+\rangle_A + e^{i \alpha'}\sin({\theta'}) |-\rangle_A ) \;, \qquad \;\;A' |-\rangle_A = (-\cos({\theta'}) |-\rangle_A+ e^{-i \alpha'}\sin({\theta'}) | +\rangle_A) \;, \nonumber \\
B |+\rangle_B & = &( \cos({\omega}) |+\rangle_B + e^{i \beta}\sin({\omega}) |-\rangle_B ) \;, \qquad \;\;\;\;B |-\rangle_B = (-\cos({\omega}) |-\rangle_B+ e^{-i \beta}\sin({\omega}) | +\rangle_B) \;, \nonumber \\
B' |+\rangle_B & = &( \cos({\omega'}) |+\rangle_B + e^{i \beta'}\sin({\omega'}) |-\rangle_B ) \;, \qquad B' |-\rangle_B = (-\cos({\omega'}) |-\rangle_B+ e^{-i \beta'}\sin({\omega'}) | +\rangle_B) \;, \label{Bphig}
\end{eqnarray}
where $(\alpha, \alpha', \theta, \theta')$ and $(\beta, \beta', \omega, \omega')$ are arbitrary parameters.\\\\We see that the operators of eqs.\eqref{Bphi} correspond to $\theta=\theta'=\omega=\omega'=\frac{\pi}{2}$. Repeating now the previous calculation using the operators \eqref{Bphig}, one gets the slightly different expression 
\begin{eqnarray} 
\langle \phi_0 |\; {\cal C} \; |\phi_0 \rangle & =& \cos({\theta}) \cos({\omega}) + \sin({\theta}) \sin({\omega}) \cos(\alpha + \beta) \nonumber \\
& +& \cos({\theta'}) \cos({\omega}) + \sin({\theta'}) \sin({\omega}) \cos(\alpha' + \beta) \nonumber \\
& +& \cos({\theta}) \cos({\omega'}) + \sin({\theta}) \sin({\omega'}) \cos(\alpha + \beta') \nonumber \\
&-& \cos({\theta'}) \cos({\omega'}) - \sin({\theta'}) \sin({\omega'}) \cos(\alpha' + \beta') \;. \label{gzv}
\end{eqnarray}
Though, it turns out that the maximum value attained by \eqref{gzv} is precisely $2\sqrt{2}$, namely 
\begin{equation}
|\langle \phi_0 |\; {\cal C} \; |\phi_0 \rangle|_{max} = 2 \sqrt{2} \;. \label{nic}
\end{equation}
This result might be considered as a nice check of Tsirelson's bound. \\\\ {\it {\bf  {\underline{Comment}}}} In the case of spin $1/2$, for which the Hilbert space is of dimension two, the characterization of the Bell operators, eqs.\eqref{Bphi} and eqs.\eqref{Bphig}, is relatively simple. Though, as the dimension of the Hilbert space increases, the task becomes  more complex. For the purposes of the present notes, we shall rely mostly on the use of eqs.\eqref{Bphi}, suitably adapted to higher dimensional Hilbert spaces. 


\vspace{1cm}

\noindent {\it {\bf  {\underline{Exercise}}}}\\\\Consider the product state 
\begin{equation}	
|\psi \rangle =  |+\rangle_A\otimes |-\rangle_B 
\;. \label{prst}
\end{equation}
Show, by using  eqs.\eqref{Bphig},  that the  Bell-CHSH correlator does not exhibit violation, {\it i.e.}
\begin{equation}
| \langle \psi |\; {\cal C} \; |\psi \rangle  | \le 2 \;. \label{absv}
\end{equation}
{\it {\bf  {\underline{Exercise}}}}\\\\Consider the state $|\phi_0\rangle$, eq.\eqref{Bs}, and the following four  angles
\begin{equation} 
\alpha=0 \;, \qquad \alpha'= \frac{\pi}{2} \;, \qquad \beta= - \varepsilon \;, \qquad \beta'= \varepsilon \;, \label{eps}
\end{equation}
where $\varepsilon >0$ is infinitesimal. Is there violation of the Bell-CHSH inequality? 

\section{Tsirelson's bound}\label{Tsl}
As we have seen, the so-called Tsirelson bound \cite{TSI}  tells us that the maximum value attainable by the Bell-CHSH correlator is $2\sqrt{2}$. Due to its general validity, it is helpful to give a proof of this statement. It relies only on the algebraic features of the Bell operators, as expressed by eqs.\eqref{AB} and eqs.\eqref{ABBell}. Let us start be recalling a few properties of the operator norm. \\\\Let ${\cal D}: {\cal H} \rightarrow {\cal H}$ be a bounded operator defined on a Hilbert space ${\cal H}$. The operator norm of ${\cal D}$ is defined as 
\begin{equation} 
|| {\cal D} || = sup  \left\{\; \frac{||{\cal D}x ||}{||x||}, x\neq 0, x\in {\cal H} \right\}  \;. \label{on}
\end{equation}
where {\it sup} stands for the supremum and $||x||$ is the norm of the vector $x$: $||x||^2= \langle x |x\rangle$. The operator norm fulfills many properties, namely 
\begin{itemize} 
\item from the definition of the  operator norm,  it follows that 
\begin{equation} 
||{\cal D}x|| \le ||{\cal D}|| \; ||x|| \;. \label{n1}
\end{equation} 
\item the triangle inequality
\begin{equation} 
||{\cal D}_1 + {\cal D}_2|| \le ||{\cal D}_1|| + ||{\cal D}_2|| \;. \label{n2} 
\end{equation}
\item the so-called continuity of the norm 
\begin{equation} 
||{\cal D}_1 {\cal D}_2|| \le ||{\cal D}_1|| ||{\cal D}_2|| \;. \label{n3}
\end{equation}
\item if ${\cal D} = {\cal T} \otimes {\cal T}'$, then 
\begin{equation} 
||{\cal D}||_{{\cal H}_T \otimes {\cal H}_{T'}} = ||{\cal T}||_{{\cal H}_T} \; ||{\cal T}'||_{{\cal H}_{T'}} \;. \label{n4}
\end{equation}
\item if ${\cal D}$ is Hermitian, ${\cal D}^\dagger = {\cal D}$, it holds that 
\begin{equation} 
||{\cal D}^2|| = ||{\cal D}||^2  \;, \label{n5}
\end{equation}
\end{itemize}
In particular, the last equation follows by observing that, from the Cauchy-Schwarts inequality, one has 
\begin{equation} 
||{\cal D}x||^2=    \langle {\cal D}x | {\cal D} x \rangle  =  \langle x | {\cal D}^2 x\rangle  \le ||x|| \; ||{\cal D}^2 x|| \le ||x||^2 \; ||{\cal D}^2|| \;. \label{pf1}
\end{equation}  
Moreover, from the property 
\begin{equation} 
\left( sup \left\{ \; \frac{||{\cal D}x ||}{||x||}, x\neq 0, x\in {\cal H}\right\}\right)^2 = sup \left\{\; \left( \frac{||{\cal D}x ||}{||x||} \right)^2, x\neq 0, x\in {\cal H} \right\} \;,\label{suppt} 
\end{equation}
it follows
\begin{equation}
||{\cal D}||^2 \le ||{\cal D}^2|| \;. \label{pf2}
\end{equation}
Also, from the continuity of the norm, we get 
\begin{equation} 
||{\cal D}^2|| = ||{\cal D}{\cal D}|| \le ||{\cal D}|| \; ||{\cal D}|| = ||{\cal D}||^2 \;, \label{pf2}
\end{equation}
thus showing eq.\eqref{n5}. \\\\We can proceed now to prove Tsirelson's statement. Let us consider the Hermitian Bell-CHSH operator $\cal C$, eq.\eqref{cop}. From eqs.\eqref{AB} and eqs.\eqref{ABBell}, it follows 
\begin{equation} 
{\cal C}^{\dagger} {\cal C} = {\cal C}^2 = 4 \mathbb{1} - \left[ A, A'\right] \otimes \left[ B,B'\right] \;, \label{cc}
\end{equation} 
where $\mathbb{1}= \mathbb{1}_A \otimes \mathbb{1}_B$. Taking the norm operator and applying the triangle inequality, one gets 
\begin{equation}
|| {\cal C}^2|| \le 4 + || \left[ A, A'\right] || \; || \left[ B,B'\right] || \;. \label{nop1}
\end{equation}
Since 
\begin{equation}
|| \left[ A, A'\right] || \le 2 ||A|| \; ||A'|| =2 \;, \qquad || \left[ B,B'\right] ||\le 2 ||B|| \; ||B'|| =2  \;, \label{nop2}
\end{equation}
one finds 
\begin{equation} 
|| {\cal C}^2 || \le 8 \;. \label{nop3}
\end{equation} 
Let  $| \psi \rangle$ stand for any normalized state, $||\psi||=1$. From the Cauchy-Schwartz inequality we get
\begin{equation} 
|\langle \psi |\; {\cal C} \; |\psi\rangle |^2 \le ||{\cal C}\psi||^2 \le ||{\cal C}||^2 \le 8 \;, \label{nop4}
\end{equation}
so that Tsirelson's bound follows, namely 
\begin{equation} 
|\langle \psi |\; {\cal C} \; |\psi\rangle | \le 2 \sqrt{2} \;, \label{nop5}
\end{equation}
for any state $|\psi\rangle$. \\\\{\it {\bf  {\underline{Comment}}}\\\\Equation \eqref{nop1} exhibits a nice property of the operator ${\cal C}$. In the classical case, the quantities $(A,A')$ and $(B,B')$ commute, {\it i.e.} $[A,A']=[B,B']=0$. The classical bound $||{\cal C}||\le 2$ is recovered. Nevertheless, at the quantum level. $(A,A')$ and $(B,B')$ become non-commuting quantities, a key feature in order to have violation.} \\\\{\it {\bf  {\underline{Exercise}}}}\\\\ Show that 
\begin{equation} 
\left( sup \left\{ \; \frac{||{\cal D}x ||}{||x||}, x\neq 0, x\in {\cal H}\right\}\right)^2 = sup \left\{\; \left( \frac{||{\cal D}x ||}{||x||} \right)^2, x\neq 0, x\in {\cal H} \right\} \;,\label{supp}
\end{equation}

\section{Gisin's theorem}\label{Gisin}
Gisin's theorem \cite{Gisin}  is an important result about the violation of the Bell-CHSH inequality. It states that any entangled {\it pure} state violates the Bell-CHSH inequality. The theorem means that, given an entangled pure state, it is always possible to find Bell's operators $(A,A')$ and $(B,B')$ such that a violation occurs. \\\\Instead of reproducing Gisin's original argument, we adopt the strategy of providing a check of the theorem. As entangled state, we shall choose the expression given in eq.\eqref{nst}. As the value of the integer $N$ becomes large, the last term is the dominant one. As a consequence, for large  values of $N$, the state becomes close to a product  state. Therefore, as $N$ increases, we expect that the violation of the Bell-CHSH inequality decreases, approaching asymptotically the classical bound 2. \\\\In order to verify Gisin's theorem it suffices to show that the Bell-CHSH inequality is still violated for large $N$, even if the size of the violation is small. After all, the state remains entangled, although in a rather weak way. \\\\Let us proceed with the evaluation of the Bell-CHSH correlator. In this case, we shall make use of the general expression for the Bell operators given in eq.\eqref{Bphig}. It turns out that 
\begin{eqnarray}
\langle \psi |\; A\otimes B\; | \psi    \rangle_N & = & \frac{\cos(\theta) \cos(\omega)}{N} \left[ N-4 \right] + \frac{ 2 \cos(\theta) \sin(\omega)}{N} \left[1 -\sqrt{N-3} \right ] \cos(\beta)  \nonumber \\
& + & \frac{ 2 \sin(\theta) \cos(\omega)}{N} \left[1 -\sqrt{N-3} \right ] \cos(\alpha) + 
\frac{ 2 \sin(\theta) \sin(\omega)}{N} \left[\sqrt{N-3}\cos(\alpha+\beta) + \cos(\alpha-\beta) \right ]  \;. \label{ABN}
\end{eqnarray}
Therefore, for the Bell-CHSH correlator, one finds 
\begin{eqnarray}
\langle \psi |\; {\cal C}\; | \psi    \rangle_N & = & \frac{\cos(\theta) \cos(\omega)}{N} \left[ N-4 \right] + \frac{ 2 \cos(\theta) \sin(\omega)}{N} \left[1 -\sqrt{N-3} \right ] \cos(\beta)  \nonumber \\
& + & \frac{ 2 \sin(\theta) \cos(\omega)}{N} \left[1 -\sqrt{N-3} \right ] \cos(\alpha) + 
\frac{ 2 \sin(\theta) \sin(\omega)}{N} \left[\sqrt{N-3}\cos(\alpha+\beta) + \cos(\alpha-\beta) \right ]  \nonumber \\
&+ &
\frac{\cos(\theta') \cos(\omega)}{N} \left[ N-4 \right] + \frac{ 2 \cos(\theta') \sin(\omega)}{N} \left[1 -\sqrt{N-3} \right ] \cos(\beta)  \nonumber \\
& + & \frac{ 2 \sin(\theta') \cos(\omega)}{N} \left[1 -\sqrt{N-3} \right ] \cos(\alpha') + 
\frac{ 2 \sin(\theta') \sin(\omega)}{N} \left[\sqrt{N-3}\cos(\alpha'+\beta) + \cos(\alpha'-\beta) \right ]  \nonumber \\
&+& 
 \frac{\cos(\theta) \cos(\omega')}{N} \left[ N-4 \right] + \frac{ 2 \cos(\theta) \sin(\omega')}{N} \left[1 -\sqrt{N-3} \right ] \cos(\beta')  \nonumber \\
& + & \frac{ 2 \sin(\theta) \cos(\omega')}{N} \left[1 -\sqrt{N-3} \right ] \cos(\alpha) + 
\frac{ 2 \sin(\theta) \sin(\omega')}{N} \left[\sqrt{N-3}\cos(\alpha+\beta') + \cos(\alpha-\beta') \right ]  \nonumber \\
&-& 
 \frac{\cos(\theta') \cos(\omega')}{N} \left[ N-4 \right] - \frac{ 2 \cos(\theta') \sin(\omega')}{N} \left[1 -\sqrt{N-3} \right ] \cos(\beta')  \nonumber \\
& - & \frac{ 2 \sin(\theta') \cos(\omega')}{N} \left[1 -\sqrt{N-3} \right ] \cos(\alpha') -
\frac{ 2 \sin(\theta') \sin(\omega')}{N} \left[\sqrt{N-3}\cos(\alpha'+\beta') + \cos(\alpha'-\beta') \right ]  \;. \label{CN}
\end{eqnarray}
In the following Table, we report the value of the Bell-CHSH correlator $\langle {\cal C} \rangle$ for different choices  of $N$, starting fom $N=3$, corresponding to the biggest violation. As expected, the violation of the Bell-CHSH inequality decreases as $N$ becomes large. Though, it is still visible for $N=10^5$. 
\begin{center}
\vspace{1cm}
  \begin{tabular}{ l | r }
    \hline
    $N$ &Bell-CHSH correlator  $\langle {\cal{C} }\rangle $  \\ \hline
    3 & 2.403  \\ \hline
    4 & 2  \\ \hline
    10 &  2.10555 \\ \hline 
    $10^2$ & 2.03108 \\ \hline 
    $10^3$ & 2.00374 \\ \hline 
    $10^4$ & 2.00039 \\ \hline 
    $10^5$ & 2.00004  \\ \hline
  \end{tabular} \label{table1} 
\end{center}
\vspace{1cm}
One sees that, for $N=4$, there is no violation. This is easily understood by observing that the state \eqref{nst} becomes a product state for $N=4$, {\it i.e.}
\begin{equation}     
|\psi\rangle_4 = \frac{1}{2} \left( |+\rangle_A  + |-\rangle_A \right) \otimes \left(   |+\rangle_B +   |-\rangle_B        \right) \;. \label{N4}
\end{equation}
{\it {\bf  {\underline{Comment}}}} Gisin's theorem holds only for {\it pure} entangled states. For {\it mixed} states, the task is more difficult. Here, it suffices to mention that  there are examples of mixed entangled states which do not exhibit violation of the Bell-CHSH inequality. These states are known as the Werner states \cite{ZW,AP,NC,SC}. 

\vspace{1cm} 

\noindent {\it {\bf  {\underline{Exercise}}}}\\\\Consider the entangled state 
\begin{equation}	
|\psi \rangle = \frac{1}{\sqrt{1+r^2}}\left( |+\rangle_A\otimes |-\rangle_B + r |-\rangle_A\otimes |+\rangle_B\right) \;, \qquad r\neq \pm 1
\;. \label{rst}
\end{equation}
Check out Gisin's theorem.

\section{The Bell-CHSH inequality for spin $1$ and spin $3/2$  }\label{spin1}

After having discussed the Bell-CHSH inequality for spin $1/2$, it is instructive to move to spin $1$ and  spin ${3}/{2}$. This will give to the reader a  concrete idea of what will be the violation for an arbitrary spin $j$. \\\\For spin $1$, the Hilbert space is of dimension three. The spin matrices are given by 
\begin{equation}
  {J_x} = \frac{1}{\sqrt{2}}
  \left[ {\begin{array}{ccc}
    0 & 1 & 0 \\
    1 & 0 & 1 \\
    0 & 1 & 0 \\
    \end{array} } \right]
     \;, \qquad 
    {J_y} = \frac{1}{\sqrt{2}}
  \left[ {\begin{array}{ccc}
    0 & -i& 0 \\
    i & 0 & 0\\
    0 & i & 0 \\
    \end{array} } \right]
     \;, \qquad 
    {J_z} =
  \left[ {\begin{array}{ccc}
    1 & 0 & 0 \\
    0 & 0 & 0 \\ 
    0 & 0 & -1
  \end{array} } \right] \;, \label{onem}
\end{equation}
with 
\begin{equation} 
\left[ J_i, J_j\right] = i \varepsilon_{ijk} J_k \;. \label{onerl}
\end{equation}
An orthonormal basis is given by the three eigenstates of $J_z$, $(|1\rangle, |0\rangle, |-1\rangle)$, corresponding to the eigenvalues $(1,0,-1)$. To investigate the Bell-CHSH correlator, we shall consider a bipartite system $AB$ built out with two spins $1$ and select the singlet state
\begin{equation} 
|\psi_1 \rangle = \frac{1}{\sqrt{3}} \left( |1\rangle_A \otimes |-1\rangle_B - |0\rangle_A \otimes |0\rangle_B + |-1\rangle_A \otimes |1\rangle_B \right) 
\;. \label{sgone}
\end{equation}Before going any further, we need to define the Bell operators. We shall rely on eqs.\eqref{Bphi}, suitably adapted to the case of spin $1$, \cite{GP,Peruzzo:2023nrr} namely 
\begin{eqnarray} 
A |1\rangle_A & = & e^{i \alpha} |-1\rangle_A \;, \qquad A |-1\rangle_A = e^{-i \alpha} | 1\rangle_A \;, \qquad \;\;\;\;A |0\rangle_A = |0\rangle_A \nonumber \\
A' |1\rangle_A & = & e^{i \alpha'} |-1\rangle_A \;, \qquad A' |-1\rangle_A = e^{-i \alpha'} | 1\rangle_A \;, \qquad A' |0\rangle_A = |0\rangle_A \nonumber \\
B |1\rangle_B & = & e^{i \beta} |-1\rangle_B \;, \qquad B |-1\rangle_B= e^{-i \beta} | 1\rangle_B \;, \qquad \;\; \;B |0\rangle_B = |0\rangle_B \nonumber \\
B' |1\rangle_B & = & e^{i \beta'} |-1\rangle_B \;, \qquad B' |-1\rangle_B = e^{-i \beta'} | 1\rangle_B \;, \qquad B' |0\rangle_B = |0\rangle_B \;. \label{ABone}
\end{eqnarray} 
Proceeding as before, one finds 
\begin{equation} 
\langle \psi_1 |\; AB \; |\psi_1 \rangle = \frac{1}{3} \left( 1 + 2 \cos(\alpha-\beta) \right)  \;, \label{ABo1}
\end{equation}
so that, for the Bell-CHSH inequality, one has 
\begin{equation} 
\langle \psi_1 |\; {\cal C} \; |\psi_1 \rangle = \frac{2}{3} \left( 1 +  \cos(\alpha-\beta) + \cos(\alpha'-\beta) + \cos(\alpha-\beta') - \cos(\alpha'-\beta') \right)  \;, \label{ABo2}
\end{equation}
whose maximum value is 
\begin{equation} 
\langle \psi_1 |\; {\cal C} \; |\psi_1 \rangle_{max}  = \frac{2}{3} (1 + 2 \sqrt{2}) = 2.55228 \;, \label{ABo3}
\end{equation}
Unlike the case of the spin $1/2$, for  spin $1$ Tsirelson's bound is not attained. As shown in \cite{GP,Peruzzo:2023nrr}, this is a general result, valid for any integer spin $j$. \\\\It can be understood in terms of the pairing mechanism outlined in \cite{GP}. When the spin is half-integer, {\it i.e.} $\frac{2s+1}{2}$, $s=0,1,etc.$, the dimension of the Hilbert space is even, {\it i.e} $(2s+2)$, so that the basis states can all be combined in pairs. \\\\Let us give an explicit example of the pairing mechanism for spin $3/2$. We  have four states: $(|\frac{3}{2}\rangle, |\frac{1}{2}\rangle, |-\frac{3}{2}\rangle , |-\frac{1}{2}  \rangle)$. Out of these states, we can form two pairs, say: $( |\frac{3}{2}\rangle, |-\frac{3}{2} \rangle)$ and $(|\frac{1}{2}\rangle, |-\frac{1}{2} \rangle )$. For the Bell operators we have 
\begin{eqnarray}
A   \Big| \frac{3}{2}{\Big \rangle}_A  & = & e^{i \alpha_1} \Big| -\frac{3}{2}{\Big \rangle} _A \;, \qquad A   \Big|-\frac{3}{2} {\Big \rangle}_A = e^{-i \alpha_1} \Big| \frac{3}{2}{\Big \rangle} _A  \nonumber \\
A   \Big|\frac{1}{2} {\Big\rangle}_A  & = & e^{i \alpha_2} \Big|-\frac{1}{2} {\Big \rangle} _A \;, \qquad A   \Big|-\frac{1}{2} {\Big \rangle}_A = e^{-i \alpha_2} \Big| \frac{1}{2}{ \Big \rangle} _A  \;, \label{Athree}
\end{eqnarray} 
with similar expressions for $(A', B, B')$. To check out the violation of the Bell-CHSH inequality in the case of spin $3/2$, we start from the expression of the singlet state for an arbitrary spin $j$ \cite{GP,Peruzzo:2023nrr}
\begin{equation}
| \psi_j \rangle = { \frac{1}{ (2j+1)^{\frac{1}{2}} } } \sum_{m=-j}^{m=j} (-1)^{j-m} | m\rangle_A \otimes | -m \rangle_B \;. \label{sj}
\end{equation}
Thus 
\begin{equation} 
| \psi_\frac{3}{2} \rangle = \frac{1}{2} \left(   \Big| \frac{3}{2}{\Big \rangle}_A \otimes \Big| -\frac{3}{2}{\Big \rangle}_B  
-  \Big| \frac{1}{2}{\Big \rangle}_A \otimes \Big| -\frac{1}{2}{\Big \rangle}_B  
+  \Big| -\frac{1}{2}{\Big \rangle}_A \otimes \Big| \frac{1}{2}{\Big \rangle}_B
- \Big|- \frac{3}{2}{\Big \rangle}_A \otimes \Big| \frac{3}{2}{\Big \rangle}_B
\right)  \;. 
\end{equation}
An easy computation shows that 
\begin{equation} 
\langle \psi_\frac{3}{2} |\; AB \; | \psi_\frac{3}{2} \rangle = - \frac{1}{2} \left(  \cos(\alpha_1 - \beta_1) + \cos(\alpha_2-\beta_2)        \right) 
\end{equation}
yielding the following expression for the spin $3/2$ Bell-CHSH correlator 
\begin{eqnarray} 
\langle \psi_\frac{3}{2} |\; {\cal C} \; | \psi_\frac{3}{2} \rangle & = &  - \frac{1}{2} \left(  \cos(\alpha_1 - \beta_1) + \cos(\alpha_2-\beta_2)        \right)  
 - \frac{1}{2} \left(  \cos({\alpha_1}' - \beta_1) + \cos({\alpha_2}'-\beta_2)        \right)  \nonumber \\
 & - & \frac{1}{2} \left(  \cos(\alpha_1 - {\beta_1}') + \cos(\alpha_2-{\beta_2}')        \right)  
  +   \frac{1}{2} \left(  \cos({\alpha_1}' - {\beta_1}') + \cos({\alpha_2}-{\beta_2}')        \right)  \label{Cthreehalf}
\end{eqnarray}
We see therefore that each pair gives a maximum contribution equals to $2\sqrt{2}$:
\begin{equation} 
| \langle \psi_\frac{3}{2} |\; {\cal C} \; | \psi_\frac{3}{2} \rangle |_{max} = \frac{1}{2} 4 \sqrt{2} = 2 \sqrt{2} \;, \label{Tsithree}
\end{equation}
so that Tsirelson's bound is attained for half-integer spins. Instead, in the case of integer spins, the dimension of the Hilbert space is odd, {\it i.e.} $(2j+1), j=1,2,3,etc$. The pairing mechanism is not perfect.  A state, say the state $|0\rangle$, see eqs.\eqref{ABone},  will be always left unpaired, decreasing a little bit the maximum value of the violation. As a consequence, Tsirelson's bound will not be attained for integer spins. \\\\{\it {\bf  {\underline{Exercise}}}}\\\\Consider the singlet state $| \psi_j \rangle$ given in equation \eqref{sj}. Show that:
\begin{itemize} 
\item for any integer spin $j \ge 1$ 
\begin{equation} 
|\langle \psi_j |\; {\cal C} \; | \psi_j \rangle|_{max} = \frac{2}{(2j+1)} \left( 1+ 2j \sqrt{2})\right)  < 2\sqrt{2} \;. \label{jint}
\end{equation}
\item for any half-integer spin 
\begin{equation} 
|\langle \psi_j |\; {\cal C} \; | \psi_j \rangle|_{max}  =  2 \sqrt{2}  \;. \label{jhint}
\end{equation}
\end{itemize}

\section{ Coherent states} \label{coh}

So far, we have studied the violation of the Bell-CHSH inequality for finite dimensional Hilbert spaces. We face now the more complex case of infinite dimensional Hilbert spaces, considering the examples of the coherent and squeezed states \cite{jpg,SandersReview}. \\\\The so called coherent states are an important tool of Quantum Mechanics, with a variety of applications in many areas as diverse as: Quantum Optics, Condensed Matter Physics, Nuclear Physics and Elementary Particle Physics.  \\\\In order to introduce coherent states, let us recall a few basic features of the infinite dimensional Fock space for the Harmonic oscillator 
\begin{eqnarray} 
H & = & \omega (a^{\dagger} a + \frac{1}{2} ) \nonumber \\
\left[a, a^{\dagger} \right] & = & 1 \;, \qquad [a,a]=[a^{\dagger}, a^{\dagger}]=0 \;. \label{hosc}
\end{eqnarray}
The vacuum Fock state $|0\rangle$ is defined by the condition 
\begin{equation} 
a |0\rangle =0 \;, \qquad \langle 0 | 0\rangle = 1 \;. \label{Fv}
\end{equation}
For the Fock basis, we have 
\begin{equation} 
\vert n \rangle = \frac{(a^\dagger)^n}{\sqrt{n!}} \vert 0 \rangle \;, \qquad \langle n'| n\rangle = \delta_{n'n}  \;, \label{basis}
\end{equation}
with 
\begin{equation} 
a \;\vert n \rangle = \sqrt{n}\; |n-1 \rangle \;, \qquad a^{\dagger} \;|n\rangle = \sqrt{n+1}\; |n+1\rangle \;, \qquad H \;|n\rangle = \omega(n+\frac{1}{2}) \;|n\rangle \;, \label{nbas} 
\end{equation}
and 
\begin{equation} 
\sum_{n=0}^{\infty} |n\rangle \langle n | =  \mathbb{1} \;. \label{compl}
\end{equation}
In order to introduce the coherent states, it is useful to make use of the  displacement operator 
\begin{equation} 
{\cal D}(z)= e^{z a^{\dagger} - z^{*} a} \;, \label{Dz}
\end{equation}
where $z=x+iy$ is a complex number. \\\\From 
\begin{equation} 
e^{X} e^{Y} = e^{(X+Y +\frac{1}{2} [X,Y] )}   \;, \qquad {\rm valid \; when} \;\;\;\; [X,[X,Y]]=[Y,[X,Y]]=0 
\end{equation}
it follows that ${\cal D}(z)$ is an unitary operator 
\begin{equation} 
{\cal D}(z)  {\cal D}^{\dagger}(z) = {\cal D}^{\dagger}(z) {\cal D}(z) = \mathbb{1} \;, \label{und}
\end{equation} 
and that it can be cast in the equivalent form 
\begin{equation} 
{\cal D}(z) = e^{-\frac{ |z|^2 }{2}} \; e^{z a^{\dagger}} \; e^{-z^{*} a} \;. \label{equiv} 
\end{equation}
Moreover, from 
\begin{equation} 
e^{X}\; Y \; e^{-X}= Y + \sum_{k=1}^{\infty} \frac{1}{k!} \underbrace{\left[X, \left[.....\left[X,Y\right].....\right],\right]}_\text{k times} \;, \label{kt}
 \end{equation}
it follows that 
\begin{equation} 
{\cal D}^{\dagger}(z)\; a\; {\cal D}(z) = a+z \;. \label{apz}
\end{equation}
We have now all ingredients to introduce the coherent states $\{ |z\rangle \}$. These states are obtained by acting with the displacement operator ${\cal D}(z)$ on the Fock vacuum $| 0 \rangle$:
\begin{equation} 
|z\rangle = {\cal D}(z) |0\rangle = e^{-\frac{|z|^2}{2}}\sum_{n=0}^{\infty} \frac{z^n}{\sqrt{n!}}\; |n\rangle  \;. \label{cz}
\end{equation}
Since ${\cal D}(z)$ is unitary, we have 
\begin{equation}
\langle z | z \rangle = \langle 0 |\; {\cal D}^{\dagger}(z) {\cal D}(z) \; |0 \rangle = 1 \;. \label{nz}
\end{equation}
Also 
\begin{equation} 
a\; |z\rangle = a \; {\cal D}(z) |0\rangle = {\cal D}(z) \left({\cal D}^{\dagger}(z) \; a \; {\cal D}(z) \right) |0\rangle = {\cal D}(z) ( a+ z) |0\rangle = z {\cal D}(z) |0\rangle = z \; |z\rangle  \;. \label{fdz}
\end{equation}
We see that the coherent state $|z\rangle$ is eigenstate of the annihilation operator $a$ with eigenvalue given by the complex number $z$. Equation \eqref{fdz} expresses a fundamental property of the coherent states. \\\\{\it \bf {\underline {Exercise}}}\\\\Show the resolution of the unity 
\begin{equation} 
\int \frac{d^2 z}{\pi} \; |z\rangle \langle z| = \mathbb{1} \;. \label{resunit}
\end{equation} 

\subsection{The pseudospin operators}\label{pseudo}

In view of the study of entangled coherent states and of the  ensuing violation of the Bell-CHSH inequality, it is worth to provide a discussion on the so called pseudospin operators \cite{psi1,psi2,psi3}. To that end we divide the basis states of the Fock space in even and odd states, namely 
\begin{equation}  
 |2n\rangle = \frac{(a^\dagger)^{2n}}{\sqrt{(2n)!}}\; |0\rangle \;, \qquad  |2n+1\rangle = \frac{(a^\dagger)^{2n+1}}{\sqrt{(2n+1)!}}\; |0\rangle \;, \qquad  n=0,1,2,... \;. \label{eo}
\end{equation} 
We introduce now the three operators defined as:
\begin{eqnarray}
s^{(n)}_x & = & \vert 2n+1 \rangle \langle 2n \vert + \vert 2n \rangle \langle 2n+1 \vert, \nonumber \\ 
s^{(n)}_y & = &i\left( \vert 2n \rangle \langle 2n+1 \vert   - \vert 2n+1 \rangle \langle 2n \vert \right), \nonumber \\ 
s^{(n)}_z & = & \vert 2n+1 \rangle \langle 2n+1 \vert - \vert 2n \rangle \langle 2n \vert. \label{spin2}
\end{eqnarray} 
An easy calculation shows that 
\begin{align} 
\left[s^{(n)}_x,s^{(n)}_y \right] &= 2 i s^{(n)}_z \;, \nonumber \\
\left[s^{(n)}_y,s^{(n)}_z \right] &= 2 i s^{(n)}_x \;, \nonumber \\
\left[s^{(n)}_z,s^{(n)}_x \right] &= 2 i s^{(n)}_y \;.  \label{spin4}
\end{align} 
It follows thus that 
\begin{equation} 
s_x = \sum_{n=0}^\infty s^{(n)}_x \;, \qquad s_y = \sum_{n=0}^\infty s^{(n)}_y \;, \qquad s_z = \sum_{n=0}^\infty s^{(n)}_z \;, \label{spin1} 
\end{equation}
obey the same algebraic relations of the spin $1/2$ Pauli matrice
\begin{align} 
	\left[ s_x,s_y \right] = 2 i s_z, \quad 	\left[s_y,s_z \right] = 2 i s_x, \quad 	\left[s_z,s_x \right] = 2 i s_y, \label{spin5}
\end{align} 
from which the name {\it pseudospin} follows. \\\\In particular, from expressions~\eqref{spin2} one observes that the introduction of the pseudospin operators can be related, once again,  to a pairing mechanism in Hilbert space, a pair being given by two states, namely $(\vert2n \rangle, \vert2n+1\rangle)$, with $n=0,1,2,..$. Each pair  gives rise to a set of operators, $(s^{(n)}_x,s^{(n)}_y,s^{(n)}_z)$, which obey the algebra of the  spin $1/2$ Pauli matrices.  As we shall see in the sequel, these operators will be very helpful in the discussion of the entanglement properties of coherent as well as of squeezed states. 

\subsection{Entangled coherent states} \label{entcoh}

We are now ready to investigate an example of entangled coherent state, see \cite{DeFabritiis:2023bbs} and refs. therein. We start by considering a bipartite system ${\cal H}= {\cal H}_A \otimes {\cal H}_B$, where ${\cal H}_A$ and ${\cal H}_B$ denote the Hilbert spaces of two Harmonic oscillators. An entangled coherent state can be introduced by means of the antisymmetric combination 
\begin{align}\label{AsymC}
\vert \psi_a \rangle = {\cal N} \left[ \vert \eta \rangle_A \otimes \vert \sigma \rangle_B + e^{i \phi} \vert -\eta \rangle_A \otimes  \vert -\sigma \rangle_B \right],
\end{align}
where $\phi$ is an arbitrary phase factor, $(\eta, \sigma)$ are real parameters and $(\vert \eta \rangle_A, \vert \sigma\rangle_B)$ are coherent states defined, respectively, in ${\cal H}_A$ and ${\cal H}_B$: 
\begin{eqnarray} 
\vert \eta \rangle_A & = & {\cal D}_A(\eta) |0\rangle_A = e^{-\frac{ \eta^2 }{2}} \; e^{\eta a^{\dagger}} \; e^{-\eta a} \; |0\rangle_A \nonumber \\
\vert \sigma \rangle_B & = & {\cal D}_B(\sigma) |0\rangle_B = e^{-\frac{ \sigma^2 }{2}} \; e^{\sigma b^{\dagger}} \; e^{-\sigma b} \; |0\rangle_B \;, \label{ABcoh}
\end{eqnarray}
with $(a,a^\dagger)$, $(b, b^\dagger)$ the creation and annihiliation operators of ${\cal H}_A$ and ${\cal H}_B$. For the normalization factor ${\cal N}$ one finds 
\begin{align}\label{key}
	{\cal N} = \frac{1}{\sqrt{2}} \left(1+ \cos\phi \exp\left[-2\left(\eta^2+\sigma^2\right)\right]\right)^{-1/2} \;. 
\end{align}
We move then to the construction of Bell's operators $(A,A',B,B')$. To that aim we rely on the definition  of the pseudospin operators given in the previous section and define 
\begin{align}\label{BellOpallmodes}
	A \vert 2n \rangle_A &= e^{i \alpha} \vert 2n + 1 \rangle_A  \;, \qquad 
	A \vert 2n + 1, \rangle_A  = e^{-i \alpha} \vert 2n \rangle_A  \nonumber \\
	A' \vert 2n \rangle_A &= e^{i \alpha'} \vert 2n + 1 \rangle_A  \;, \qquad 
	A' \vert 2n + 1, \rangle_A  = e^{-i \alpha'} \vert 2n \rangle_A  \nonumber \\
	B \vert 2m \rangle_B &= e^{i \beta} \vert 2m + 1 \rangle_B  \;, \qquad 
	B \vert 2m + 1, \rangle_B  = e^{-i \beta} \vert 2m \rangle_B  \nonumber \\
	B' \vert 2m \rangle_B &= e^{i \beta'} \vert 2m + 1 \rangle_B  \;, \qquad 
	B' \vert 2m + 1, \rangle_B  = e^{-i \beta'} \vert 2m \rangle_B  \ ;. 
\end{align}
These operators fulfill all required conditions, eqs.\eqref{AB}.\eqref{ABBell}. In terms of pesusdospin operators, we have 
\begin{equation} 
A = {\vec u} \cdot {\vec s}_A   \otimes {\mathbb{1}}_B,  \label{A2st}
\end{equation}
where $\vec{u}$ is the unit vector
\begin{equation}
{\vec u} =\left( \cos(a), -\sin(a),0 \right), \qquad {\vec u} \cdot {\vec u} = 1, \label{vecu}
\end{equation} 
Similar expressions hold for the operators $(A',B,B')$. \\\\{\bf {\underline{Comment}}}\\\\{\it Although a more general form for the operators $(A,A',B,B')$ can be introduced, see eqs.\eqref{Bphig}, expressions \eqref{BellOpallmodes} will enable us to already capture the violation of the Bell-CHSH inequality for both coherent and squeezed states.} \\\\In order  to evaluate  the Bell-CHSH correlator  when the   state $\vert \psi_a \rangle$ is employed, {\it i.e.}
\begin{equation}
\langle \psi_a  |\; {\cal C} \; | \psi_a \rangle = \langle \psi_a  |\; (A+A')\otimes B + (A-A')\otimes B' \; | \psi_a \rangle  \;. \label{BCoh}
\end{equation}
 we look at the term $\langle \psi_a \vert\; A \otimes B \;\vert \psi_a \rangle$. Making use of the properties of the Fock basis, it turns out that 
  \begin{equation}\label{key}
	\langle \psi_ a |\;A \otimes B \;  \psi_a |\rangle_{A} = 4 \Omega_{a} \, \left( \cos(\alpha)  \cos( \beta) - \cos(\phi) \sin(\alpha)  \sin(\beta) \right) \Delta(\eta,\sigma)  \;, 
\end{equation} 
with 
\begin{equation}
\Delta(\eta, \sigma) = \sum_{n,m=0}^{\infty}\left[\frac{\eta^{4n+1} \sigma^{4m+1}}{\sqrt{(2n)! (2n+1)! (2m)! (2m+1)!}}\right] \;, \label{Dt}
\end{equation}
end
 \begin{align}\label{OmegaA}
	\Omega_{a} = \frac{\exp\left[-\left(\eta^2 + \sigma^2\right)\right]}{1+ \cos\phi \exp\left[-2\left(\eta^2 + \sigma^2\right)\right]}.
\end{align}
Therefore, for the Bell-CHSH correlator, we obtain 
\begin{eqnarray} 
\langle \psi_a |\; {\cal C} \;| \psi_a \rangle& =&  4 \Omega_{a} \Delta(\eta,\sigma) \, \left( \cos(\alpha)  \cos( \beta) - \cos(\phi) \sin(\alpha)  \sin(\beta)  
+ cos(\alpha')  \cos( \beta) - \cos(\phi) \sin(\alpha')  \sin(\beta) \right) \nonumber \\
&+ &4 \Omega_a \Delta(\eta,\sigma) \left( cos(\alpha)  \cos( \beta') - \cos(\phi) \sin(\alpha)  \sin(\beta')
- cos(\alpha')  \cos( \beta') + \cos(\phi) \sin(\alpha')  \sin(\beta)
\right)  \label{Cc}
\end{eqnarray} 
Although this expression looks pretty simple, the infinite sum cannot be written in a closed  form. To analyze the violation, 
 we choose to work with $\phi = \pi $ and  $(\alpha = 0,\, \alpha' = \pi/2,\, \beta = +\pi/4,\, \beta' = -\pi/4)$. \\\\Firstly, we consider the first contribution of the sum, taking the $n=m=0$ term. It  can be written as:
\begin{align}
	\langle\psi_a |\;  A \otimes B \; | \psi_a \rangle \vert_{n=m=0} = \frac{4 \sqrt{2} \, \eta  \sigma}{\sinh\left(\eta ^2+\sigma^2\right)} \;.  \label{nm}
\end{align}
For small values of $\eta $ and $\sigma$ expression \eqref{nm} is already  close to $2\sqrt{2}$.  We face then the full expression, eq.\eqref{Cc}. Taking $\eta = \sigma = 0.1$ and performing a numerical analysis of the sum, we already find $\langle \mathcal{C} \rangle = 2.8284 \simeq 2 \sqrt{2}$. We increase now the values of $(\eta, \sigma)$. For $\eta = \beta = 1$, we still have $\langle \mathcal{C} \rangle = 2.6678$. \\\\ It is worth noticing that a violation of the  Bell-CHSH inequality takes place  even in the case without a relative phase, that is, when $\phi=0$. In fact, taking $\phi=0$ one  can consider  a slightly different set of parameters: $(\alpha = 0,\, \alpha' = \pi/2,\, \beta = -\pi/4,\, \beta' = +\pi/4)$. Thus, for $\eta=\sigma=0.7$ one finds $\langle \mathcal{C} \rangle = 2.0895$, see \cite{DeFabritiis:2023bbs} for a more general  discussion. \\\\In conclusion, one can state that the entangled coherent state \eqref{AsymC} exhibits a violation of the Bell-CHSH correlator for a wide range of the parameters. \\\\{\it  \bf {\underline{Exercise}}}\\\\Consider the 
symmetric entangled coherent state 
\begin{align}\label{Symc}
	\vert \psi \rangle_s = {\cal N}_s \left[ \vert \eta \rangle_A \otimes \vert \sigma \rangle_B + e^{i \phi} \vert \sigma \rangle_A \otimes \vert \eta \rangle_B \right] \;. \
\end{align}
Evaluate the normalization constant ${\cal N}_s$ and discuss the violation of the Bell-CHSH inequality. \\\\{\it \bf  {\underline{Exercise}}}\\\\Consider the Schr\"odinger cat states, {\it i.e.},
\begin{align}\label{Catc}
	\vert \psi \rangle_\pm = C_\pm \left[\vert \eta \rangle_\pm \otimes \vert \sigma \rangle_\pm + e^{i \phi}  \vert \sigma \rangle_\pm \otimes \vert \eta \rangle_\pm\right],
\end{align}
where $C_\pm$ is a normalization factor and  
\begin{align}\label{Schcat}
	\vert \eta \rangle_\pm = N_\pm \left[ \vert \eta \rangle \pm \vert -\eta \rangle  \right] \;..
\end{align}
Evaluate the normalization factors and analyse the violation of the Bell-CHSH inequality.

\section{ Squeezed states} \label{sq}

Similarly to coherent states, the squeezed states \cite{jpg,DeFabritiis:2023gos} find applications in many areas. These states can be defined as 
\begin{equation}
| \lambda \rangle = \sqrt{1-\lambda^2} \sum_{n=0}^{\infty} \lambda^n |n_a n_b\rangle  \;, \qquad 0< \lambda <1 \;, \label{sqs}
\end{equation}
where $|n_a n_b \rangle$ is a shorthand notation for 
\begin{equation} 
| n_a n_b \rangle = \frac{(a^\dagger)^{n}}{\sqrt{n!}} \otimes \frac{(b^\dagger)^{n}}{\sqrt{n!}} |0\rangle \;, \qquad |0\rangle = |0\rangle_A \otimes |0\rangle_B \;, \label{nn}
\end{equation}
where $(a^\dagger, b^\dagger)$ stand for the creation operators of ${\cal H}_A$ and $ {\cal H}_B$. The parameter $\lambda$ is known as the squeezing parameter. Let us check that the state $| \lambda \rangle$ is correctly normalized. From 
\begin{equation} 
\langle n_a n_b \; | \; m_a m_b \rangle = \delta_{nm} \label{nmz}
\end{equation}
it follows that 
\begin{equation}
\langle \lambda\; | \; \lambda \rangle = (1 -\lambda^2) \sum_{n=0}^{\infty} \lambda^{2n} = \frac{1-\lambda^2}{1-\lambda^2} = 1  \;. \label{lone}
\end{equation}
To investigate the violation of the Bell-CHSH inequality we rewrite the squeezed state in terms of even and odd Fock states, namely 
\begin{equation}
| \lambda \rangle = \sqrt{1-\lambda^2} \left( \sum_{n=0}^{\infty} \lambda^{2n} |(2n)_a (2n)_b\rangle + \sum_{n=0}^{\infty} \lambda^{2n+1} |(2n+1)_a (2n+1)_b\rangle \right) \;.  \label{loe}
\end{equation}
Making use of the Bell operators of eqs.\eqref{BellOpallmodes}, it turns out that 
\begin{equation} 
A \otimes B |\lambda\rangle = \sqrt{1-\lambda^2} \left( e^{i(\alpha+ \beta)} \sum_{n=0}^{\infty} \lambda^{2n} |(2n+1)_a (2n+1)_b\rangle 
+ e^{-i(\alpha+ \beta)} \sum_{n=0}^{\infty} \lambda^{2n+1} |(2n)_a (2n)_b\rangle \right) \;. \label{ABl}
\end{equation}
Therefore 
\begin{equation} 
\langle \lambda |\; A \otimes B \; | \lambda\rangle = (1-\lambda^2) \;2 \lambda \cos(\alpha+\beta) \sum_{n=0}^{\infty} \lambda^{4n} = \frac{2 \lambda}{1+\lambda^2} \cos(\alpha+\beta) \;. \label{lABl}
\end{equation} 
As a consequence, the Bell-CHSH correlator is found 
\begin{equation}
\langle \lambda |\; {\cal C}  \; | \lambda\rangle = \langle \lambda |\; (A+ A') \otimes B + (A- A') \otimes B'\; | \lambda\rangle = 
\frac{2 \lambda}{1+\lambda^2}\left( \cos(\alpha+\beta)+ \cos(\alpha'+\beta) + \cos(\alpha+\beta') - \cos(\alpha'+\beta') \right) \;. \label{lC}
\end{equation} 
The angular contribution attains its maximum value for 
\begin{equation} 
\alpha = 0 \;, \qquad \beta= - \frac{\pi}{4} \;, \qquad \alpha'= \frac{\pi}{2} \;, \qquad \beta'= \frac{\pi}{4} \;, \label{stv}
\end{equation}
yielding 
\begin{equation} 
\langle \lambda |\; {\cal C}  \; | \lambda\rangle = \frac{4 \sqrt{2}  \lambda}{1+\lambda^2} \;. \label{Csqz}
\end{equation}
There is violation whenever 
\begin{equation} 
|\langle \lambda |\; {\cal C}  \; | \lambda\rangle | > 2  \;,  \label{lviol}
\end{equation}
which occurs for 
\begin{equation} 
 \sqrt{2}-1 < \lambda < 1 \;. \label{lviol1}
\end{equation}
The maximum violation is attained when $\lambda$ is very close to $1$: 
\begin{equation} 
|\langle \lambda |\; {\cal C}  \; | \lambda\rangle |_{\lambda \approx 1} \approx 2 \sqrt{2} \;. \label{maxl}
\end{equation}

\section{ Mermin's inequalities and GHZ states} \label{mermin}

As we have seen in the previous examples, the formulation of the Bell-CHSH inequality is valid for  a bipartite system $AB$. Nevertheless, there are  generalizations of the Bell-CHSH inequality for  multipartite systems \cite{Mermin}, see also \cite{DeFabritiis:2023gos} and refs therein. These generalized inequalities are known as the Mermin inequalities. \\\\Let us discuss in detail the case of the Mermin inequality for a three partite system $ABC$. The Hilbert space of the system is ${\cal H}={\cal H}_A \otimes {\cal H}_B \otimes {\cal H}_C$. Proceeding as in Section \eqref{Bineq}, we consider three pairs of classical quantities $(a,a'), (b,b'),(c,c')$  taking values $\pm 1$, {\it i.e.} 
\begin{equation} 
a^2 = a'^2= b^2= b'^2 = c^2 = c'^2 = 1 \;. \label{threeone}
\end{equation}
Let us prove that the absolute value of the combination 
\begin{equation}
   a'  b c + a b'c + a b c' - a' b' c'   \;, \label{abs}
\end{equation}
is bounded by 2. We observe that the pair $(c,c')$ can take the four values $(1,1), \;  (1,-1) ,\; (-1,1), \;(-1,-1)$. Though, since we are considering the absolute value, we can analyze only two cases: $(1,1)$ and $(1,-1)$. It is easy to check out that in both cases one goes back to the classical combination already encountered in the case of the Bell-CHSH inequality, Sect. \eqref{Bineq}. In fact, in the firts case we have 
\begin{equation} 
 \vert  a'  b c + a b'c + a b c' - a' b' c'  \vert_{(1,1)} =  \vert  a'  b + a b' + a b  - a' b'   \vert  = \vert (a+a') b + (a-a')b \vert \le 2 \;. \label{fcase}
\end{equation}
Similarly, in the second case one finds 
\begin{equation} 
 \vert  a'  b c + a b'c + a b c' - a' b' c'  \vert_{(1,-1)} = \vert a' b + a b' - a b + a' b' \vert= \vert (a'- a) b + (a+a') b'\vert \le \vert a'- a\vert + |a + a'\vert \le 2 \;. \label{scase}
\end{equation}
Thus, on classical grounds, we have 
\begin{equation} 
\vert  a'  b c + a b'c + a b c' - a' b' c'  \vert  \le 2 \;.  \label{abstwo}
\end{equation}
At the quantum level, $(a,a')$, $(b,b')$, $(c,c')$ become dichotomic Hermitian operators fulfilling the conditions 
\begin{eqnarray} 
A & =& A^{\dagger} \;, \quad A'  = {A'}^{\dagger} \;, \quad B  = B^{\dagger} \;, \quad B'  = {B'}^{\dagger} \;, \qquad C= C^\dagger \;, \qquad C'= {C'}^\dagger \nonumber \\[3mm]
A^2 & =& {A'}^2 = B^2 = {B'}^2=C^2= {C'}^2=1 \;, \label{ABC}
\end{eqnarray}
and
\begin{eqnarray}
\left[\; A,B\; \right] &= & 0\;, \quad \left[\; A',B\; \right] =  0 \;, \quad \left[\; A,B'\; \right] =  0 \;, \quad \left[\; A',B'\; \right] =  0 \;. \left[\; A,C\; \right] =  0\;, \quad \left[\; A',C\; \right] =  0 \nonumber \\[3mm]
\left[\; B,C\; \right] &= & 0\;, \quad \left[\; B',C\; \right] =  0 \;, \quad \left[\; B,C'\; \right] =  0 \;, \quad \left[\; B',C'\; \right] =  0 \;.  \nonumber \\[3mm]
\left[\; A,A'\; \right] &\neq & 0\;, \qquad \left[\; B,B'\; \right] \neq 0 \;, \qquad  \left[\; C,C'\; \right] \neq  0 \;.   
\label{ABCMl}
\end{eqnarray}
Following \cite{Mermin}, we introduce the Hermitian operator 
\begin{equation} 
{\cal M}_3 = {\cal M}^{\dagger}_3 = A' \otimes B \otimes C\; + \;A \otimes B' \otimes C\; + A\otimes B\otimes C' \; - \; A' \otimes  B' \otimes C' \;. \label{M3}
\end{equation}
Repeating the same procedure employed in the derivation of Tsirelson's bound, it follows 
\begin{equation} 
{\cal M}_3^2 = 4 \mathbb{1} - \left[ A, A' \right] \otimes \left[ B, B' \right] - \left[ A, A' \right] \otimes \left[ C, C' \right] -\left[ B, B' \right] \otimes \left[ C, C' \right] \;. \label{Tsm}
\end{equation}
Using the properties of the operator norm and recalling that each commutator contributes with a factor 2, we get 
\begin{equation} 
|| {\cal M}^2_3|| = ||{\cal M}_3 ||^2 \le 16 \;. \label{nm3}
\end{equation}
As a consequence, for any normalized state $| \psi \rangle$, $\langle \psi | \psi\rangle =1$, we have 
\begin{equation} 
\vert \langle \psi |\; {\cal M}_3 \;| \psi \rangle \vert \le 4 \;. \label{bm3}
\end{equation}
The operator ${\cal M}_3$ is called the Mermin operator of order three. One speaks of a violation of the Mermin inequality whenever 
\begin{equation} 
2 < \vert \langle \psi |\; {\cal M}_3 \;| \psi \rangle \vert \le 4 \;. \label{bm3v}
\end{equation}
Let us look now at the analogue of the Bell states in the case of the ${\cal M}_3$ inequality. These states are known as the  
Greenberger-Horne-Zeilinger states \cite{GHZ}, simply called GHZ states, being given by the expression 
\begin{equation} 
\vert \psi_{GHZ} \rangle = \frac{ |+\rangle_A \otimes |+\rangle_B \otimes |+\rangle_C - |-\rangle_A \otimes |-\rangle_B \otimes |-\rangle_C}{\sqrt{2}} \;. \label{GHZ}
\end{equation} 
From 
\begin{eqnarray} 
A |+\rangle_A & = & e^{i \alpha} |-\rangle_A \;, \qquad A |-\rangle_A = e^{-i \alpha} | +\rangle_A \;, \nonumber \\
A' |+\rangle_A & = & e^{i \alpha'} |-\rangle_A \;, \qquad A' |-\rangle_A = e^{-i \alpha'} | +\rangle_A \;, \nonumber \\
B |+\rangle_B & = & e^{i \beta} |-\rangle_B \;, \qquad B |-\rangle_B = e^{-i \beta} | +\rangle_B \;, \nonumber \\
B' |+\rangle_B & = & e^{i \beta'} |-\rangle_B \;, \qquad B' |-\rangle_B = e^{-i \beta'} | +\rangle_B \;, \nonumber \\
C |+\rangle_C & = & e^{i \gamma} |-\rangle_C \;, \qquad C |-\rangle_C = e^{-i \gamma} | +\rangle_C \;, \nonumber \\
C' |+\rangle_C & = & e^{i \gamma'} |-\rangle_C \;, \qquad C' |-\rangle_C = e^{-i \gamma'} | +\rangle_C \;,\label{ABCact}
\end{eqnarray} 
one obtains 
\begin{equation} 
A B C \vert \psi_{GHZ}\rangle =  \frac{- \;e^{-i(\alpha + \beta + \gamma)}  |+\rangle_A \otimes |+\rangle_B \otimes |+ \rangle_C + e^{i(\alpha+\beta+\gamma)} |-\rangle_A \otimes |-\rangle_B \otimes |-\rangle_C}{\sqrt{2}} \;, \label{ABCGHZ}
\end{equation} 
so that 
\begin{equation} 
\langle \psi_{GHZ} |\; ABC  \;| \psi_{GHZ} \rangle = \cos(\alpha +\beta +\gamma) \;.  \label{ABC} 
\end{equation}
Thus, for the Mermin inequality, we get 
\begin{equation} 
 \langle \psi_{GHZ} |\; {\cal M}_3  \;| \psi_{GHZ} \rangle= \cos(\alpha' +\beta +\gamma) + \cos(\alpha +\beta' +\gamma) + \cos(\alpha +\beta +\gamma') - \cos(\alpha' +\beta' +\gamma') \;.  \label{m3ABC} 
\end{equation}
Setting 
\begin{equation} 
\alpha = 0 \;, \qquad \alpha'= \frac{\pi}{2}\;, \qquad \beta = - \frac{\pi}{4} \;, \qquad \beta'= \frac{\pi}{4} \;, \qquad \gamma= - \frac{\pi}{4} \;, \qquad \gamma'= \frac{\pi}{4} \;, \label{m3angles}
\end{equation}
one gets maximum violation, {\it i.e.} 
\begin{equation} 
\vert \langle \psi_{GHZ} |\; {\cal M}_3  \;| \psi_{GHZ} \rangle \vert_{max} = 4 \;. \label{maxm3} 
\end{equation}
We see thus that the GHZ state, eq.\eqref{GHZ}, maximizes the Mermin inequality of order three.\\\\{\it  \bf {\underline{Exercise}}}\\\\Consider the Mermin operator of order 4 
\begin{eqnarray} 
2 {\cal M}_4 & = & - ABCD + (A'BCD + AB'CD + ABC'D + ABC D')  + (A'B'CD + A'B C'D + A'B C D') \nonumber \\
&+& ( A B'CD + A B'C D'+ A B C'D') - (A'B'C'D + A'B'C D'+ A'B C'D'+ A B'C'D') - A'B'C'D'  \;, \label{m4}
\end{eqnarray} 
where $(A,A',B,B',C,C'D,D')$ are dichotomic Hermitian operators. 
Show that, for any normalized state $|\psi\rangle$, there is violation whenever 
\begin{equation} 
2< \vert \psi |\;  {\cal M}_4 \; | \psi \rangle \vert \le 4 \sqrt{2} 
\end{equation}
Consider the GHZ state 
\begin{equation} 
\vert \psi_{GHZ} \rangle_4 = \frac{ |+\rangle_A \otimes |+\rangle_B \otimes |+\rangle_C \otimes |+ \rangle_D - |-\rangle_A \otimes |-\rangle_B \otimes |-\rangle_C \otimes |- \rangle_D }{\sqrt{2}} \;. \label{GHZ}
\end{equation} 
Show that $\vert \psi_{GHZ} \rangle_4$ gives maximum violation of the ${\cal M}_4$-inequality 
\cite{DeFabritiis:2023ubj}.

\section{Conclusion} 
Since its discovery, Bell's inequality is the source of renewed investigations. In these notes we have provided a concise and self-contained account on a version of the Bell inequality known as the Bell-CHSH inequality. \\\\Many examples have been worked out, covering both finite and infinite dimensional Hilbert spaces. \\\\The extension of the Bell-CHSH inequality to the multipartite Mermin inequalities has also been outlined. \\\\We hope that the material will be helpful to graduate as well as undergraduate students aiming at pursuing research projects on this fascinating topic.

\section*{Acknowledgments}
	The authors would like to thank the Brazilian agencies CNPq and CAPES, for financial support.  S.P.~Sorella, I.~Roditi, and M.S.~Guimaraes are CNPq researchers under contracts 301030/2019-7, 311876/2021-8, and 309793/2023-8, respectively.

\appendix

\section{The Bell theorem}\label{A}	
	
For the sake of completeness, let us reproduce here the main steps of Bell's theorem \cite{Bell64}. We shall follow ref.\cite{Bell:1980wg}, where Bell's theorem is presented for the Bell-CHSH combination. \\\\According to \cite{Bell:1980wg}, we consider a bipartite system $AB$ made up by two spins $1/2$ in the singlet state: 
\begin{equation} 
|\phi_{s}\rangle = \frac{1}{\sqrt{2}}\left( |+\rangle_A\otimes |-\rangle_B - |-\rangle_A\otimes |+\rangle_B\right) \;. \label{bl1}
\end{equation} 
The Bell operators can be written as 
\begin{equation} 
A = {\vec a}\cdot {\vec \sigma}_A \;, \qquad  A' = {\vec a}\cdot {\vec \sigma}_A \;, \qquad B = {\vec b}\cdot {\vec \sigma}_B \;, \qquad  B' = {\vec b}'\cdot {\vec \sigma}_B \;, \label{bl2}
\end{equation}
where $({\vec \sigma}_A, {\vec \sigma}_B)$ are the Pauli matrices acting, respectively, on Alice  and Bob Hilbert spaces. The vectors $({\vec a}, {\vec a}')$ and $({\vec b}, {\vec b}')$  are unit vectors specifying the directions in which the spin measurements are performed. 
The operators \eqref{bl2}
fulfill the conditions
\begin{eqnarray} 
A & =& A^{\dagger} \;, \quad A'  = {A'}^{\dagger} \;, \quad B  = B^{\dagger} \;, \quad B'  = {B'}^{\dagger} \;, \nonumber \\[3mm]
A^2 & =& {A'}^2 = B^2 = {B'}^2=1 \;, \label{bl3}
\end{eqnarray}
and
\begin{eqnarray}
\left[\; A,B\; \right] &= & 0\;, \quad \left[\; A',B\; \right] =  0 \;, \quad \left[\; A,B'\; \right] =  0 \;, \quad \left[\; A',B'\; \right] =  0 \;.  \nonumber \\[3mm]
\left[\; A,A'\; \right] &\neq & 0\;, \quad \left[\; B,B'\; \right] \neq  0 \;.   
\label{bl4}
\end{eqnarray}
Quantum Mechanics gives 
\begin{equation} 
\langle \psi_s |\; A \otimes B \; | \psi_s \rangle =  - \cos(\theta_{ab}) \;, \label{cos}
\end{equation}
where $\theta_{ab}$ is the angle between ${\vec a}$ and ${\vec b}$. When ${\vec a}={\vec b}$, {\it i.e.} when the measurements are performed in the same direction, 
\begin{equation} 
 \langle \psi_s |\; A \otimes B \; | \psi_s \rangle \Big|_{\theta_{ab}=0} =  - 1  \;, \label{coss1}
 \end{equation}
meaning that a perfect anti-correlation takes place, As we have already seen, for the Bell-CHSH correlator, one finds maximum violation: 
\begin{equation}
\Big| \langle \psi_s |\; {\cal C} \; | \psi_s \rangle \Big| = \Big| \langle \psi_s |\; (A +A') \otimes B+ (A -A') \otimes B' \; | \psi_s \rangle \Big| = 2 \sqrt{2} \;. \label{mxbl}
\end{equation}
Within the realm of a {\it loca} hidden variables theory, one attempts at reproducing the Quantum Mechanical results, eqs.\eqref{cos},\eqref{mxbl}, by introducing a set of hidden variables $\{ \lambda \}$ together with a distribution probability $\rho(\lambda)$ 
\begin{equation} 
\int d\lambda\; \rho(\lambda) = 1 \;. \label{dist}
\end{equation}
Expression \eqref{cos} should thus be reproduced by 
\begin{equation} 
{\cal E}(a,b) =\int d\lambda \; \rho(\lambda) \; {\cal A}(a,\lambda) {\cal B}(b, \lambda) \;, \label{reprod}
\end{equation}
where ${\cal A}(a,\lambda)$ and $ {\cal B}(b, \lambda)$ are suitable random variables taking values $\pm 1$, ${\cal A}^2(a,\lambda)=1$, $ {\cal B}^2(b, \lambda)=1$, and 
\begin{equation} 
{\cal B}(b=a, \lambda)= - {\cal A}(a, \lambda) \;. \label{amb}
\end{equation}
It should be noticed that, in writing eq.\eqref{reprod}, the important requirement of locality has been taken into account. It means that the quantity 
${\cal A}(a,\lambda)$ cannot depend on the settings of Bob's device, namely: ${\cal A}(a,\lambda)$  does not depend on ${\vec b}$. Similarly, $ {\cal B}(b, \lambda)$ cannot depend on ${\vec a}$. As emphasized by Bell \cite{Bell64}, this is a key feature of the whole reasoning. \\\\It is now easy to prove that the Quantum Mechanical results cannot be reproduced by any local hidden variable theory, {\it i.e.} by expressions of the type of eq.\eqref{reprod}. Take in fact the combination 
\begin{equation} 
{\cal E}(a,b)+ {\cal E}(a',b)+{\cal E}(a,b')- {\cal E}(a',b) = \int d\lambda \; \rho(\lambda) \;{\cal C}(a,b, \lambda) \;, \label{appc}
\end{equation} 
with 
\begin{equation} 
{\cal C}(a,b, \lambda) = {\cal A}(a,\lambda) {\cal B}(b, \lambda) +{\cal A}(a',\lambda) {\cal B}(b, \lambda) +{\cal A}(a,\lambda) {\cal B}(b', \lambda) -{\cal A}(a',\lambda) {\cal B}(b', \lambda)  \;. \label{ccc}
\end{equation} 
Making use of the Cauchy-Schwarz inequality 
\begin{equation} 
\Big| \int d\lambda\; \rho(\lambda) \;  {\cal C}(a,b, \lambda) \Big|^2 \le \int d\lambda \; \rho(\lambda) \; \Big| {\cal C}(a,b, \lambda) \Big|^2  \;, \label{cscs}
\end{equation}
and of the commuting nature of $({\cal A}(a,\lambda),  {\cal A}(a',\lambda),{\cal B}(b, \lambda),{\cal B}(b', \lambda))$, one finds 
\begin{equation} 
{\cal C}(a,b, \lambda)^2 =4 \;. \label{four} 
\end{equation} 
As a consequence, it follows that 
\begin{equation}
\Big| {\cal E}(a,b)+ {\cal E}(a',b)+{\cal E}(a,b')- {\cal E}(a',b) \Big| \le 2 \;, \label{norep}
\end{equation} 
showing that the Quantum Mechanical results cannot be reproduced by local hidden variables theories. This statement expresses the content of Bell's theorem.


\end{document}